\documentclass[]{raa}       
\usepackage{graphicx,times} 
\usepackage{natbib,amsmath}
\usepackage{amssymb}
\bibpunct{(}{)}{;}{a}{}{,}
\usepackage[a4paper=true,driverfallback=dvips,pagebackref=true]{hyperref}
\usepackage{tabularx, booktabs, threeparttable}

\newcolumntype{L}[1]{>{\raggedright\arraybackslash}p{#1}}
\newcolumntype{R}[1]{>{\raggedleft\arraybackslash}p{#1}}

\def\apj{The Astrophysical Journal}
\def\apjl{Astrophysical Journal Letters}
\def\aap{Astronomy and Astrophysics}

\def\prd{Physical Review D}

\def\apjs{The Astrophysical Journal Supplement Series}

\def\be{\begin{equation}}
\def\ee{\end{equation}}
\def\bq{\begin{eqnarray}}
\def\eq{\end{eqnarray}}
\def\page{p_{\rm age}}

\begin{document}

\title{A More Accurate Parameterization based on cosmic Age (MAPAge)
}

   \volnopage{Vol.0 (20xx) No.0, 000--000}      
   \setcounter{page}{1}          

   \author{Lu Huang
      \inst{}
   \and Zhiqi Huang$^*$
      \inst{}
   \and Zhuoyang Li
      \inst{}  
   \and Huan Zhou
       \inst{}
   
   }
   \institute{School of Physics and Astronomy, Sun Yat-Sen University, 2 Daxue Road, Tangjia, Zhuhai, 519082, P.R.China
   {\it huangzhq25@mail.sysu.edu.cn}\\
\vs\no
   {\small Received~~20xx month day; accepted~~20xx~~month day}}

   \abstract{Recently, several statistically significant tensions between different cosmological datasets have raised doubts about  the standard Lambda cold dark matter ($\Lambda$CDM) model. A recent letter~\citet{Huang:2020mub} suggests to use ``Parameterization based on cosmic Age'' (PAge) to approximate a broad class of beyond-$\Lambda$CDM models, with a typical accuracy $\sim 1\%$ in angular diameter distances at $z\lesssim 10$. In this work, we extend PAge to a More Accurate Parameterization based on cosmic Age (MAPAge) by adding a new degree of freedom $\eta_2$. The parameter $\eta_2$ describes the difference between physically motivated models and their phenomenological PAge approximations. The accuracy of MAPAge, typically of order $10^{-3}$ in angular diameter distances at $z\lesssim 10$, is significantly better than PAge. We compare PAge and MAPAge with current observational data and forecast data. The conjecture in~\citet{Huang:2020mub}, that PAge approximation is sufficiently good for current observations, is quantitatively confirmed in this work. We also show that the extension from PAge to MAPAge is important for future observations, which typically requires sub-percent accuracy in theoretical predictions.}

   \authorrunning{Huang et al. }            
   \titlerunning{MAPAge approximation}  

   \maketitle


\section{Introduction}        

 At the end of the last century, the observed extra dimming of Type Ia supernovae (SNe) led to the introduction of dark energy into standard cosmology~\citep{perlmutter1999astrophys,riess1998observational,Scolnic:2017caz}. In the concordance $\Lambda$ cold dark matter ($\Lambda$CDM) model, dark energy is interpreted as the cosmological constant invented by Albert Einstein. Despite the great observational success of $\Lambda$CDM model in the last two decades~\citep{Planck18Params, BOSS2021, DES3yr}, the fine tuning problem and the coincidence problem of the cosmological constant remain to be a pain in the neck for at least some, if not all, of the theoreticians~\citep{Weinberg:1988cp,Zlatev:1998tr}. The concordance model was further challenged by the recently emerged observational discrepancy between directly and indirectly measured Hubble constants~\citep{Riess:2020fzl,Wong:2019kwg} and a $\sim 3\sigma$ tension between the measurements of the matter clustering~\citep{Asgari:2020wuj}. The theoretical unnaturalness and observational tensions have motivated a plethora of beyond-$\Lambda$CDM models~\citep{Quintessence98, Quintessence99, Kessence_Chiba, Kessence_APC, DGP, GCG01, GCG02, fR03, CoupledQuintessence, RunMp}, although in the Bayesian view none of them has been proven to be significantly more competitive than $\Lambda$CDM. The zoology of the models, namely the exercise of computing Bayesian evidences for all the models, if ever possible, may not be a pleasant job. Preferred choices are often more phenomenological models such as a perfect-fluid dark energy with its equation of state being a constant ($w$CDM model) or a linear function of the scale factor ($w_0$-$w_a$ model)~\citep{Chevallier:2000qy,Linder:2002et}. More blind methods such as Taylor expansion~\citep{Visser:2003vq} and Gaussian process~\citep{GP12} are sometimes used to explore more complex scenarios.

The Parameterization based on the cosmic Age (PAge), recently proposed by \citet{Huang:2020mub}, is somewhat in between. It is a semi-blind model capturing common physical features of many models, such as matter dominance at high redshift and that the energy density of the universe decreases with time. In the Bayesian view, PAge is more economic than many other bottom-up methods in the sense that it only contains one more parameter than $\Lambda$CDM. Physical models can be mapped to PAge space either by simply matching the deceleration parameter at some pivot redshift or by doing a least square fitting of cosmological observables. It has been shown that the fitting errors of distance moduli is typically $\lesssim 1\%$ at $z\lesssim 10$~\citep{Luo:2020ufj, Huang:2020evj}. Empirically such an accuracy is good enough for utilizing current cosmological data at $z\lesssim 10$. Indeed, PAge has been applied to many currently available data sets and yielded fruitful results~\citep{Huang:2020mub, Luo:2020ufj, Huang:2020evj,Cai:2021weh}.

Tiny deviation from PAge, however, may become measurable with future cosmological surveys. For instance, the ongoing ground-based project of Dark Energy Spectroscopic Instrument (DESI) will measure angular diameter distance and Hubble parameter to percent-level accuracy up to redshift $z\sim 1.5$~\citep{Aghamousa:2016zmz}. The precision of measurements is expected to be further improved by the space missions of Euclid satellite and the Chinese Space Station Telescope Optical Survey in the near future~\citep{Euclid, CSST}.

Thus, aiming at future cosmological surveys, we extend PAge to a ``More Accurate Parameterization based on the cosmic Age'' (MAPAge) by adding a new degree of freedom into PAge, of which the ansatz will be given immediately below in Section~\ref{sec:mapage}. We show in Section~\ref{sec:constrain} that the new degree of freedom in MAPAge is not well constrained by current observations, thus explicitly confirming that PAge is accurate enough for current data. In Section~\ref{sec:forecast}, we forecast the constraint on MAPAge parameters with simulated data of baryon acoustic oscillations (BAO) from DESI. We conclude and discuss in Section~\ref{sec:conclusion}.

\section{Model}  \label{sec:mapage}

In PAge approximation, the expansion rate of the Universe is parameterized as~\citep{Huang:2020mub} 
\be
 \frac{H}{H_0} = 1+\frac{2}{3}\left(1-\eta\frac{H_0 t}{\page}\right)\left(\frac{1}{H_0 t}-\frac{1}{\page}\right), \label{eq:page}
\ee
where $t$ is the cosmological time and $H$ is the Hubble parameter. Here $H_0\equiv 100h\, \rm km\,s^{-1}Mpc^{-1}$ is the Hubble constant. The age parameter $\page\equiv H_{0}t_0$ is the present cosmological time $t_0$ expressed in unit of $H^{-1}_0$. The phenomenological parameter $\eta$ ($\eta < 1$) characterizes the deviation from Einstein de-Sitter universe (flat CDM model).

In numeric calculation,  the correspondence between the cosmological time $t$ and the redshift $z$ is obtained by inverting the monotonic function
\be
z(t) = e^{-\int_0^t H(t') dt'} - 1.
\ee
It can be trivially shown that $H(z)$ is automatically guaranteed to be a monotonically increasing function. At high redshift $z\gg 1$, PAge asymptotically approaches the matter-dominated behavior $a \propto t^{2/3}$, where $a=\frac{1}{1+z}$ is the scale factor. (As a late-universe phenomenological approximation, PAge ignores the very short period of radiation-dominated era.) These key features makes PAge a very compact approximate description of many physical models.

The More Accurate version of PAge, MAPAge, is formulated to preserve the aforementioned advantages,
\be
 \frac{H}{H_0}=1+\frac{2}{3}\left ( 1- \left ( \eta+\eta_2 \right )\frac{H_0t}{\page}+\eta_{2} \left ( \frac{H_0t}{\page} \right )^{2} \right)\left (  \frac{1}{H_0t}-\frac{1}{\page} \right ), \label{eq:mapage}
\ee
where the new degree of freedom $\eta_2$ ($-1<\eta_2<1$) can be regarded as a cubic-order correction to PAge approximation. When $\eta_2=0$, MAPAge degrades to PAge.

\begin{figure}
\centering
\includegraphics[width=0.7\textwidth]{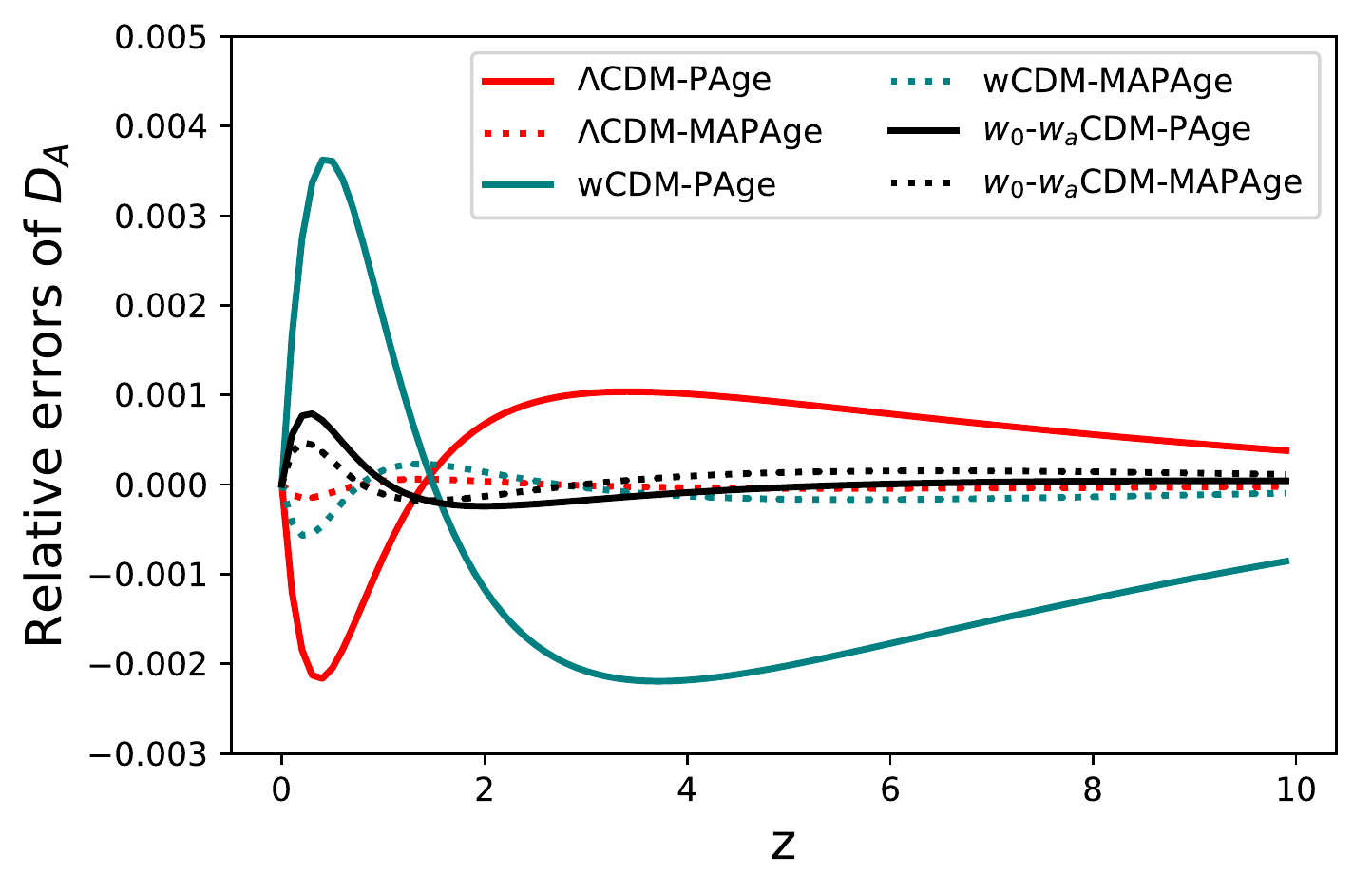}
\caption{PAge approximation and MAPAge approximation: relative errors of angular diameter distance. Model parameters are given in Table.~\ref{table:comparison}.}
\label{fig:figureDA}
\end{figure}

The cosmological models listed in~\citet{Luo:2020ufj} and ~\citet{Huang:2020evj} can be approximately mapped into MAPAge space with better precision. A least-square fitting of the dimensionless quantity $Ht$ (as a function of redshift) is applied in the redshift range $0\le z\le 10$ to determine the phenomenological parameters $\eta$ and $\eta_2$. Table~\ref{table:comparison} shows the fitting accuracy of angular diameter distances $D_A$. Comparing these results with the fitting accuracy of PAge, given in \citet{Luo:2020ufj} and \citet{Huang:2020evj}, we find that the precision of MAPAge approximation is typically $\sim$ an order of magnitude better than PAge. Such a comparison is not entirely fair, however,  because in \citet{Luo:2020ufj} and \citet{Huang:2020evj} the $\eta$ parameter of PAge is determined differently, by matching the deceleration parameter $q_0$ at redshift zero. To obtain a more fair comparison, we apply the same least square fitting method to PAge. The $D_A$ fitting errors of PAge and MAPAge for a few models are shown in Figure~\ref{fig:figureDA}. The result again confirms MAPAge's superiority in fitting accuracy.

\begin{table*}
\caption{\label{table:comparison} MAPAge approximation:  maximum relative errors in $D_A(z)$ ($0\le z\le 10$). }\centering
\begin{tabular}{llllll}
\hline\hline
models &  parameters & $ \page$ & $\eta$ & $\eta_2$  &max$\left | \frac{\Delta D_{A}}{D_{A}} \right |$ \\
\hline
CDM & $\Omega_m=1$ & $\frac{2}{3}$ & 0& 0&0\\

nonflat CDM & $\Omega_m=0.3,\Omega_k=0.7$ & 0.809 &-0.0389 & -0.373& $3.71\times10^{-3}$\\

$\Lambda$CDM & $\Omega_m=0.3$& 0.964& 0.377& 0.0752&$1.53\times10^{-4}$\\

nonflat $\Lambda$CDM & $\Omega_m=0.5,\Omega_k=0.2$& 0.797& 0.113&-0.0596& $7.52\times10^{-4}$\\

$w$CDM & $\Omega_m=0.3,w=-1.2$ & 0.991& 0.664& -0.148&$5.77\times10^{-4}$\\

nonflat $w$CDM &$\Omega_m=0.33,\Omega_k=-0.25,w=-0.8$ &0.967& 0.215&0.411& $1.85\times10^{-3}$\\

$w_0-w_a$CDM & $\Omega_m=0.3,w_0=-1,w_a=0.3$ & 0.953& 0.372&-0.0131& $4.69\times10^{-4}$\\

nonflat $w_0-w_a$CDM  & $\Omega_m=0.25,\Omega_k=0.1,w_0=-1.2,w_a=-0.2$ & 1.009& 0.629&-0.197&$1.78\times10^{-3}$\\

GCG & $\Omega_b=0.05, A=0.75,\alpha=0.1$ & 0.956&0.426&0.0386&$5.66\times10^{-4}$\\

DGP & $\Omega_m=0.3$ & 0.907& 0.148& -0.0252& $4.20\times10^{-5}$\\
\hline
\end{tabular}
\end{table*}

\section{Current Constraints} \label{sec:constrain}

To study the observational constraint on MAPAge parameters, we compile the BAO data sets from 6dF Galaxy Survey and Sloan Digital Sky Survey (SDSS)~\citep{Beutler:2011hx, Ross:2014qpa, Alam:2016hwk, Bourboux:2017cbm, Ata:2017dya} and the Pantheon SNe data~\citep{Scolnic:2017caz}. A $\chi$-square likelihood is applied to both data sets. The collected data points and covariance matrices can be found in \citet{Luo:2020ufj} and \citet{Scolnic:2017caz}, respectively. We perform Monte Carlo Markov Chain (MCMC) calculation for both PAge and MAPAge. Uniform priors are applied on $hr_{d}\in[0,200\mathrm{Mpc}]$, $ \page \in [0.8,1.2]$, $\eta \in [-2,1]$,  and $\eta_2 \in [-1,1]$. 

\begin{table*}
\caption{\label{table:constraint} Marginalized constraints on Page and MAPAge parameters}\centering
\begin{tabular}{lllllll}
\hline\hline
models & data & $hr_{d}/\mathrm{Mpc}$ & $\page$ & $\eta$ & $\eta_2$ & $ \chi^{2}_{\rm min}/\mathrm{d.o.f}$\\
\hline
PAge & current BAO+SNe & $101.2\pm 1.2$ & $0.975\pm 0.013$ & $0.372\pm 0.065$ &  - &$0.987$ \\
MAPAge  & current BAO+SNe  &$102.0\pm 1.3$ & $0.968\pm 0.014$ & $0.50\pm 0.12$ & $-0.32\pm 0.24$ &$0.985$ \\
MAPAge & BAO forecast &$100.4\pm 0.74$ &  $0.960\pm 0.0079$ & $0.394\pm 0.044$ &$0.04\pm 0.12$ &$0.00284$\\
\hline
\end{tabular}
\end{table*}

\begin{figure}
\centering
\includegraphics[width=0.95\textwidth]{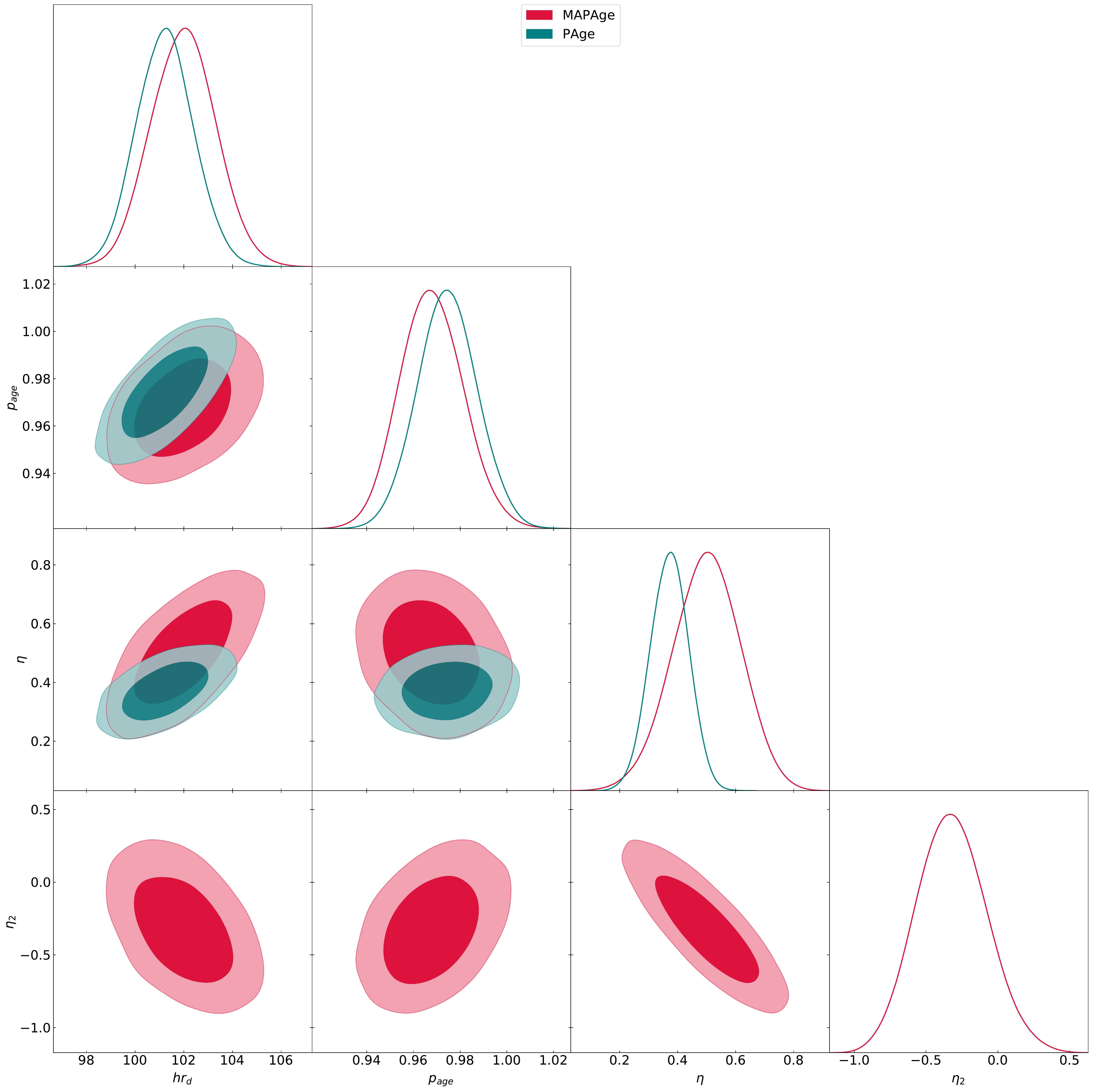}
\caption{The marginalized constraints on PAge and MAPAge parameters. The inner and outer contours enclose $68\%$ and $95\%$ confidence regions, respectively.}
\label{fig:comparison}
\end{figure}

We present the marginalized constraints (mean and 68\% confidence-level bounds) on PAge and MAPAge parameters in Table.~\ref{table:constraint}. The marginalized 1$\sigma$, 2$\sigma$ contours for both PAge and MAPAge parameters are shown in Figure~\ref{fig:comparison} for a comparison. The apparent worsen measurement of $\eta$ in MAPAge is due to the strong degeneracy between $\eta$ and $\eta_2$, as shown by the red $\eta$-$\eta_2$ contour in Figure~\ref{fig:comparison}.  The constraints on  $hr_d$ and $\page$ parameters, different from $\eta$,   are not significantly influenced by  the inclusion of $\eta_2$. Here we see that the coincidental proximity $p_{\rm age}\approx 1$,  which has inspired some recent discussion about whether we are living in a special cosmic era~\citep{Avelino2016}, is a rather robust result that does not rely on the $\Lambda$CDM framework.

The strong degeneracy between $\eta$ and $\eta_2$ suggests that the data cannot distinguish MAPAge and PAge very well. It is therefore unnecessary to use MAPAge for current data, in the spirit of Occam's razor principle. As an easy and efficient parameterization, PAge approximation still serves as a sufficiently powerful tool to investigate late-time cosmological expansion history beyond the $\Lambda$CDM physic, such as the late-time cosmic acceleration~\citep{Huang:2020mub, Luo:2020ufj}, the statistically significant $H_0$ crisis~\citep{Riess:2020fzl,Wong:2019kwg} and $S_8$ tension~\citep{Asgari:2020wuj}. However, with the rapid accumulation of cosmological data, the situation might change within a few years. In the next section we proceed to explore the future prospects of MAPAge.

\section{Forecast}\label{sec:forecast}

DESI is an ongoing Stage IV ground-based dark energy experiment. It is designed to study BAO and the growth of structure with a wide-area galaxy and quasar redshift survey. DESI experiment aims to provide more precise and higher quality observation data, at least an order of magnitude improvement over SDSS both in the comoving volume and the number of galaxies~\citep{Aghamousa:2016zmz}. These more precise data can greatly advance our understanding of the cosmic evolution history and the nature of cosmic dark components.

We adopt the DESI BAO forecast data from different tracers across the redshift range $[0, 3.5]$ and covering 14,000 square degrees of the sky. The radial and transverse BAO forecast are listed in Table.2.3, Table.2.5 and Table.2.7 of \citet{Aghamousa:2016zmz}, respectively. The baseline cosmological model employed in DESI forecast has been mentioned in Section 2.4.1 of \citet{Aghamousa:2016zmz}, whose fiducial values are detailedly summarized in Table 5 of \citet{Planck:2013pxb}. This fiducial flat-$\Lambda$CDM cosmology approximately corresponds to $hr_d=100.13$, $\page=0.957$, $\eta=0.370$ and $\eta_2=0.072$ in MAPAge language.  We perform MCMC analysis with a $\chi$-square likelihood. The last row of Table~\ref{table:constraint} lists the marginalized posteriors of MAPAge parameters, which are consistent with the  mapping results of MAPAge derived from the fiducial cosmology.

\begin{figure}
\centering
\includegraphics[width=0.95\textwidth]{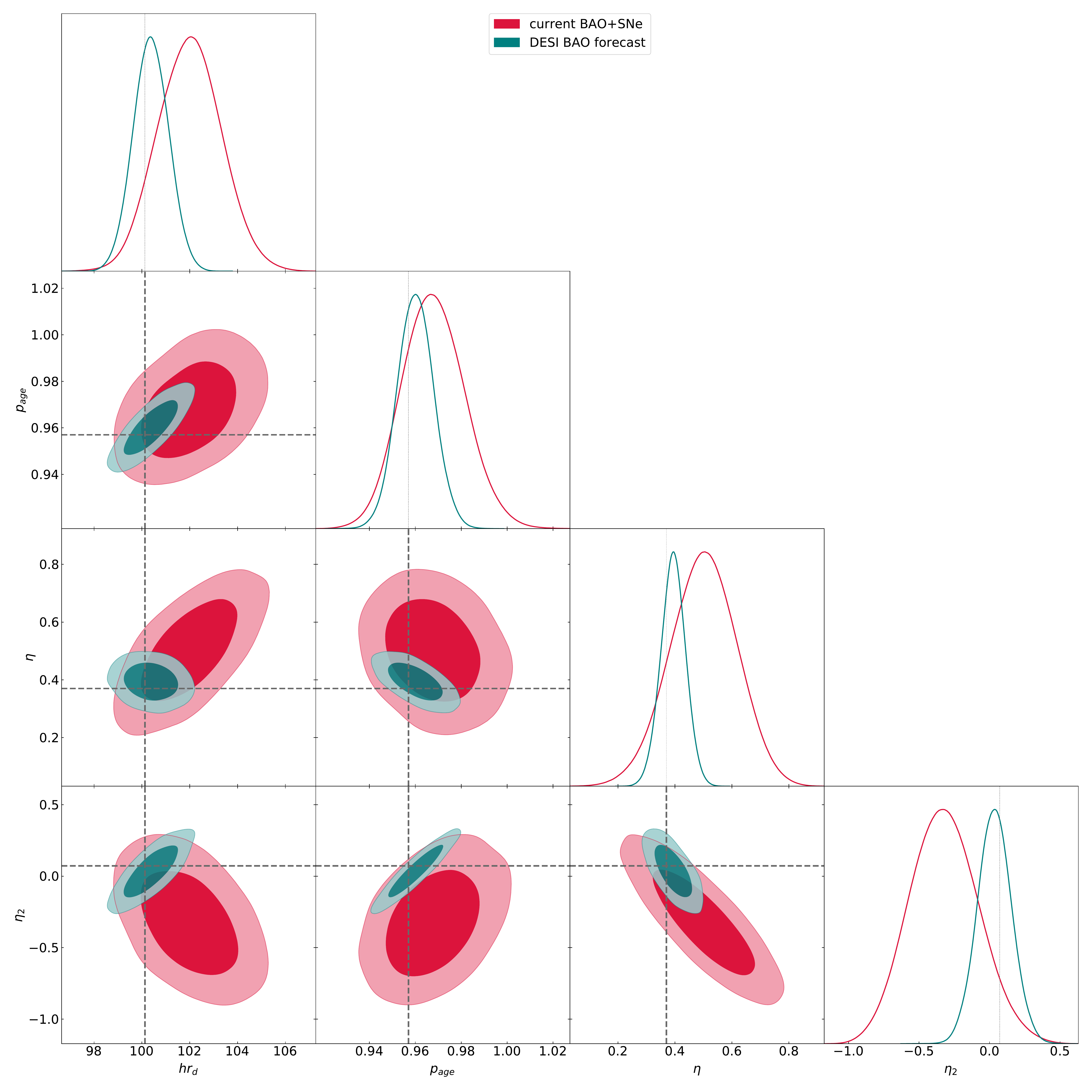}
\caption{Marginalized constraints on MAPAge parameters with current BAO+SNe and forecast DESI BAO data.  The grey dashed horizontal and vertical lines show the mapping values of MAPAge which correspond to the fiducial $\Lambda$CDM cosmology applied in forecast.}
\label{fig:forecast}
\end{figure}

In Figure~\ref{fig:forecast} we compare the constraints on MAPAge parameters with current BAO+SNe data and with forecast DESI BAO data. The apparent shrinking of the parameter by parameter contours  from now (BAO + SNe) to future (DESI) suggests that MAPAge will be come more and more favorable as the clock ticks.

\section{Conclusion}\label{sec:conclusion}

Among many phenomenological late-universe expansion reconstruction methods, PAge has been proven to be a very compact and robust one. In this work, we extended the PAge approximation to a more precise framework (MAPAge) to embrace the forthcoming era of hyper-precision cosmology. MAPAge approximation inherits all the advantages of PAge and can approximate the expansion history of many models to $\sim 10^{-3}$ level. We took the ongoing Stage IV experiment DESI as an example and showed the superiority of MAPAge for cosmological surveys in the near future.

\section{Acknowledgements}
 This work is supported by the National SKA Program of China No. 2020SKA0110402, National key R\&D Program of China (Grant No. 2020YFC2201602), Guangdong Major Project of Basic and Applied Basic Research (Grant No. 2019B030302001), and National Natural Science Foundation of China (NSFC) under Grant No. 12073088. 


\end{document}